\documentclass[prl,twocolumn,showpacs]{revtex4}
\usepackage{graphicx,epsf}

\newcommand{\bq}{{\bf q}}
\newcommand{\bk}{{\bf k}}
\newcommand{\bp}{{\bf p}}
\newcommand{\br}{{\bf r}}

\newcommand{\nzero}{{\rm N_0}}
\newcommand{\nphi}{{\rm N_\phi}}

\newcommand{\beq}{\begin{equation}}
\newcommand{\eeq}{\end{equation}}
\newcommand{\beqn}{\begin{eqnarray}}
\newcommand{\eeqn}{\end{eqnarray}}
\newcommand{\nn}{\nonumber}

\begin{document}

\def\tende#1{\,\vtop{\ialign{##\crcr\rightarrowfill\crcr
\noalign{\kern-1pt\nointerlineskip}
\hskip3.pt${\scriptstyle #1}$\hskip3.pt\crcr}}\,}

\title{A bosonization approach for bilayer quantum Hall systems at $\nu_T = 1$}

\author{R. L. Doretto,$^{1,2}$ A. O. Caldeira,$^2$ and C. Morais Smith$^1$}

\affiliation{$^1$Institute for Theoretical Physics,
                 Utrecht University,
                 Postbus 80.195, 3508 TD Utrecht,
                 The Netherlands \\
$^2$Departamento de F\'{\i}sica da Mat\'eria Condensada,
    Instituto de F\'{\i}sica Gleb Wataghin,
    Universidade Estadual de Campinas,
    CEP 13083-970, Campinas-SP, Brazil}

\begin{abstract}
We develop a non-perturbative bosonization approach for bilayer
quantum Hall systems at $\nu_T = 1$, which allows us to
systematically study the existence of an exciton condensate in
these systems. An effective boson model is derived and the
excitation spectrum is calculated both in the Bogoliubov and in
the Popov approximations. In the latter case, we show that the
ground state of the system is an exciton condensate only when the
distance between the layers is very small compared to the magnetic
length, indicating that the system possibly undergoes another phase
transition before the incompressible-compressible one. 
The effect of a finite electron interlayer tunnelling is included
and a quantitative phase diagram is proposed.
\end{abstract}
\maketitle

{\it Introduction.}
The recent conjecture of Bose-Einstein condensation (BEC) of excitons
in a bilayer quantum Hall system (QHS) at {\it
total} filling factor one ($\nu_T = 1$) has attracted a great
deal of attention \cite{perspectives,eisenstein04}.
This system consists of two quantum wells (layers) separated by a
fixed distance $d$, under a perpendicular magnetic field $B$. Each
layer has filling factor $\nu = n\phi_0/B = 1/2$, where $n$
is the layer electron density and $\phi_0 = hc/e$ is the
magnetic flux quantum.
By increasing the ratio $d/l$,
where $l = \sqrt{\hbar c/eB}$ is the magnetic length, the bilayer QHS
at $\nu_T=1$ undergoes a transition from an incompressible to a
compressible phase \cite{murphy94}. Indeed, for small $d/l$,
spontaneous interlayer coherence develops and the system behaves
as a single-layer QHS with $\nu = 1$,
whereas for large values of $d/l$ intralayer correlations become
important and the system behaves as two independent QHSs with
$\nu=1/2$ \cite{perspectives,girvin01}. 

The nature of the ground state of the bilayer QHS at $\nu_T=1$
is an issue not completely settled yet.   
From the experimental point of view, tunnelling
conductance experiments \cite{spielman00} show a zero bias peak
which does not seem to diverge with decreasing temperatures. In addition, a
linear current-voltage characteristic, instead of a power law one, has
been observed in magneto-transport measurements
\cite{kellogg04,wiersma04}. Both points  
raise doubts about the existence of a true exciton condensate,
which might have superfluid properties \cite{wen92}.
Theoretically, diagrammatic calculations
\cite{fertig89}, which rely on the existence of an exciton
condensate, indicate that the ground state of the system becomes
unstable for $d/l \ge 1.2$. Single-mode approximation
\cite{macdonald90} and generalized random phase approximation 
calculations \cite{joglekar01} confirm the result. Recently,
Fertig and Murthy suggested that this ``imperfect'' two-dimensional
superfluid behavior of the bilayer may be understood in terms of
a coherence network induced by disorder, which breaks up the system
into large and small regions with, 
respectively, weak and strong interlayer coherence \cite{fertig05}.

In this letter, we develop a non-perturbative bosonization
formalism for bilayer QHSs at $\nu_T=1$, which allows us to
properly explore the idea of exciton condensation in this system.
An {\it interacting} boson model is derived and the excitation
spectrum is calculated both in the Bogoliubov and in the Popov
approximations. In the latter case, we find that the ground state
of the system is a true exciton condensate only when $d/l \le
(d/l)_c$. For the zero electron interlayer tunnelling case,
$(d/l)_c = 0.4$, but this parameter increases with the tunnelling
strength. This new phase transition (or crossover) takes place at a critical ratio 
$(d/l)_c$ {\it much smaller} than $(d/l)_c^{\rm I-C}$, where the
incompressible-compressible phase transition is experimentally
observed. Our findings, which are based on a proper treatment of the Coulomb
interaction, yield a quantitative phase diagram for the bilayers. 

{\it The model.} Let us consider a bilayer system with $N$
spinless electrons moving in the $(x,y,z=0)$ plane and $N$
spinless electrons moving in the $(x,y,z=d)$ plane in an external
magnetic field $\mathbf{B} = B\hat{z}$. For each layer, we
restrict the Hilbert space to the lowest Landau level (LLL) and consider
$N = \nphi /2$, where $\nphi$ is the Landau level degeneracy.
We also introduce a pseudospin
index ($\sigma = \uparrow$, $\downarrow$) in order to define to
which layer each electron is associated.

\begin{figure}[t]
\centerline{\includegraphics[height=4.5cm]{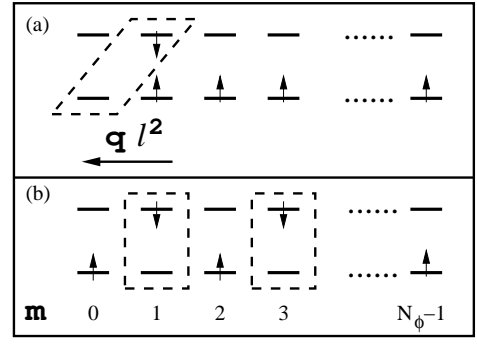}}
\caption{\label{fig01}{Schematic representation of (a) one-boson
                    state (magnetic exciton) of the quantum Hall
                    ferromagnet with $|\bq l^2| = 1$ and (b) the condensate
                    of $\nphi /2$ zero-momentum bosons.
                    The guiding center quantum number
                    is denoted by {\bf m}.}}
\end{figure}

The Hamiltonian of the system is given by
\beq
\label{total_hamiltoniano}
\mathcal{H} = \mathcal{H}_T + \mathcal{V},
\eeq
where $\mathcal{H}_T$ describes the electron tunnelling between the two
layers,
\beq
\label{tunel}
\mathcal{H}_T = -
\frac{1}{2}\Delta_{SAS}\sum_{m}c^\dagger_{m\;\uparrow}c_{m\;\downarrow}+\rm{h.c.},
\eeq
and $\mathcal{V}$ is the Coulomb interaction term (unit area system)
\begin{equation}
\label{interaction01} \mathcal{V} =
\frac{1}{2}\sum_{k}\sum_{\sigma, \sigma'}v_{\sigma
\sigma'}(k)\rho_{\sigma}(k)\rho_{\sigma'}(-k).
\end{equation}
Here $\Delta_{SAS}$ is the electron interlayer
tunnelling, the fermion operator $c^\dagger_{m\;\sigma}$ creates an electron
with guiding center $m$ in the LLL of the $\sigma$ layer (see Fig.\ \ref{fig01}),
$\rho_{\sigma}(k)$ is the Fourier transform of the projected density
operator of pseudospin $\sigma$ electrons, and
$v_{\uparrow\uparrow}(k) = $ $v_{\downarrow \downarrow}(k) =$
$v_{A}(k) = $ $(2\pi e^{2}/\epsilon k)\exp(-|kl|^{2}/2)$ and
$v_{\uparrow \downarrow}(k) = $ $v_{\downarrow \uparrow}(k) =  $
$v_{E}(k) = $ $(2\pi e^{2}/\epsilon k)\exp(-|kl|^{2}/2)\exp(-kd)$
are, respectively, the Fourier transform of the intrAlayer and
intErlayer interaction potential.

Firstly, we will concentrate on the limit $\Delta_{SAS} = 0$.
The Hamiltonian (\ref{total_hamiltoniano}) reduces to the Coulomb term
(\ref{interaction01}) which can be rewritten in terms of the total electron
density $\rho(k) = \rho_{\uparrow}(k)+\rho_{\downarrow}(k)$ and
the $z$-component of the pseudospin density $S_{Z}(k) =
\left[\rho_{\uparrow}(k)-\rho_{\downarrow}(k)\right]/2$ operators
as
\begin{equation}
\label{interaction}
\mathcal{V} = \frac{1}{2}\sum_k v_{0}(k)\rho(k)\rho(-k)+2\sum_{k}v_{Z}(k)S_{Z}(k)S_{Z}(-k),
\end{equation}
with
\[
v_{0/Z}(k) = \frac{1}{2}[v_A(k) \pm v_E(k)]=\frac{\pi e^{2}}{\epsilon  k}e^{-|kl|^{2}/2}(1 \pm e^{-kd}).
\]

The restriction of the Hilbert space to the LLL
together with the introduction of the pseudospin language renders
the description of the bilayer QHS at $\nu_T = 1$
analogous to the one of the single-layer QHS at $\nu = 1$
when the electron spin degree of freedom is included.
For the latter, it was shown that the particle-hole pair
excitations (magnetic excitons) of the ground state
(quantum Hall ferromagnet)
can be approximately treated as bosons \cite{doretto}. In this framework, the
electron and the $z$-component of the (pseudo)spin density
operators are written as 
\begin{eqnarray}
\rho(k) & = & \delta_{k,0}\nphi +2i\sum_{\bq}\sin(\bk\wedge
               \bq/2)b_{\bk+\bq}^{\dagger}b_{\bq},
\nonumber  \\
S_{Z}(k) & = & \frac{1}{2}\delta_{k,0}\nphi -\sum_{\bq}\cos(\bk\wedge
               \bq/2)b_{\bk+\bq}^{\dagger}b_{\bq},\label{rhosz}
\end{eqnarray}
with $\mathbf{k}\wedge\mathbf{q} \equiv
l^2\hat{z}\cdot(\mathbf{k}\times\mathbf{q})$. Here,
$b_{\bq}^{\dagger}$ and $b_{\bq}$ are boson operators which obey
the canonical commutation relations $
[b^{\dagger}_{\mathbf{q}},b^{\dagger}_{\mathbf{q'}}] =
[b_{\mathbf{q}},b_{\mathbf{q'}}] = 0, $ and
$[b_{\mathbf{q}},b^{\dagger}_{\mathbf{q'}}] =
\delta_{\mathbf{q},\mathbf{q'}}$. When  $b_{\bq}^{\dagger}$ is
applied to the quantum Hall ferromagnet, it creates a magnetic
exciton whose momentum $\bq$ is related to the vector $\br$
between the guiding centers of the excited electron
and hole by $|\br| = l^2|\bq|$ [Fig.\ \ref{fig01}(a)]
\cite{kallin}. In this formalism, the configuration of a bilayer
QHS at $\nu_T = 1$ corresponds to a system with $\nphi /2$
bosons.

Substituting Eqs.\ (\ref{rhosz}) into (\ref{interaction}) and
normal-ordering the result, we obtain, apart from a constant
related to the positive background, an {\it interacting} boson
model
\begin{eqnarray}
\mathcal{H}_B &=& \sum_{\bq} w_qb_{\bq}^{\dagger}b_{\bq} +
 \sum_{\bk,\bp,\bq}2v_{\bk}(\bp,\bq)b_{\bk+\bp}^{\dagger}b_{\bq-\bk}^{\dagger}b_{\bq}b_{\bp},
\label{hboso}
\end{eqnarray}
where the dispersion relation of the bosons is given by
\begin{eqnarray}
w_{q} & = & \frac{e^{2}}{\epsilon l}\left(\sqrt{\frac{\pi}{2}} - \frac{d}{l}
          - l\int_{0}^{\infty}dk\,
      e^{-kd}e^{-(kl)^2/2}{\rm J_0}(kql^2)\right),
\nn
\end{eqnarray}
with ${\rm J_0}(x)$ denoting the Bessel function of first kind.
The second term of $w_q$ is similar to the capacitor-like one
introduced phenomenologically by Girvin \cite{girvin01}.
The boson interaction potential is given by
\begin{eqnarray}
v_{\bk}(\bp,\bq) &=& v_{0}(k)\sin(\bk\wedge\bp/2)\sin(\bk\wedge\bq/2)
       \nn \\
       &+& v_{Z}(k)\cos(\bk\wedge\bp/2)\cos(\bk\wedge\bq/2).
\label{boson_interaction}
\end{eqnarray}
By taking the limit $d/l \rightarrow 0$ in Eq. (\ref{hboso}), we recover the
boson model for the single-layer QHS at $\nu=1$ \cite{doretto}.

{\it Bogoliubov approximation.} We start the analysis of the
interacting boson model (\ref{hboso}) within the
Bogoliubov approximation \cite{books}. We assume that the bosons
condense in their lowest energy state, the $\bq = 0$ mode, which means
that the ground state of the system is roughly given by
the state shown in Fig.\ \ref{fig01}(b). This assumption is in
agreement with the scenario proposed in
Refs.\cite{eisenstein04}.
Therefore, the boson operators $b_{\bq=0}^{\dagger}$ and $b_{\bq=0}$ can be
replaced by complex numbers, i.e., $b_{\bf 0}^{\dagger}, b_{\bf
0}\rightarrow \sqrt{\nzero}$, where $\nzero$ is the
(macroscopic) number of bosons in the $\bq = 0$ mode.

\begin{figure}[t]
\centerline{\includegraphics[height=9.5cm]{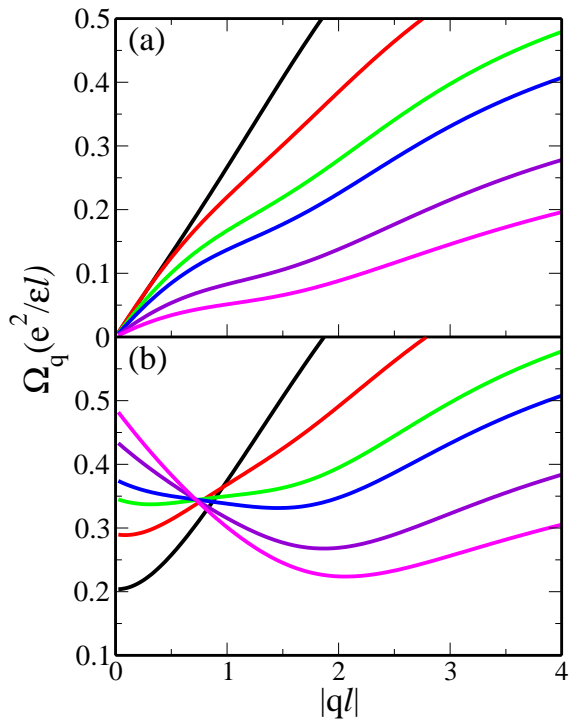}}
\caption{\label{fig02}{(color online) Dispersion relation of the
    quasiparticles in the Bogoliubov approximation 
    for different
    values of the ratio $d/l$: 0.2 (black), 0.5 (red), 0.8 (green), 1.0 (blue), 1.5
    (violet), and 2.0 (magenta). (a) $\Delta_{SAS} = 0$ and (b)
    $\Delta_{SAS} = 0.1\;e^2/\epsilon l$.}}
\end{figure}

After substituting $b_{\bf 0}^{\dagger}$ and $b_{\bf 0}$ by
$\sqrt{\nzero}$ in the Hamiltonian (\ref{hboso}), including the
chemical potential $\mu$ explicitly, i.e., $\mathcal{H}_B
\rightarrow {\rm \hat{K}} = \mathcal{H}_B - \mu\hat{N}$, and keeping
only the {\it quadratic} terms in the boson operators, we obtain
\beqn
{\rm \hat{K}} & \approx & 2v_Z(0){\rm N}_0^2 + (w_0 - \mu)\nzero 
                    -(1/2)\sum_{\bq\not=0}\epsilon_q   
\nn \\
  &+& \frac{1}{2}\sum_{\bq\not=0}\left[\epsilon_q
       (b_{\bq}^{\dagger}b_{\bq} + b_{-\bq}b_{-\bq}^\dagger) 
       + \lambda_q(b_{\bq}^{\dagger}b^\dagger_{-\bq} + 
        b_{-\bq}b_{\bq})
        \right],
\nn \\
\label{hboso_bog}
\eeqn
where $\epsilon_q =  w_q - \mu + 4\nzero v_Z(0) + 4\nzero v_Z(q)$ and $\lambda_q =
4 \nzero v_Z(q)$. The above Hamiltonian can be diagonalized by
the canonical transformation $b_\bq = \cosh(\gamma_q)a_\bq -
\sinh(\gamma_q)a^\dagger_{-\bq}$, and $b^{\dagger}_\bq =
\cosh(\gamma_q)a^\dagger_\bq - \sinh(\gamma_q)a_{-\bq},$ with
$\gamma_q$ real, and therefore the dispersion relation of the
quasiparticles is given by
\begin{equation}
\Omega_q = \sqrt{\epsilon^2_q - \lambda^2_q}.
\label{dispersion}
\end{equation}
In the one-loop approximation \cite{stoof93},
$\mu = w_0 + 4v_Z(0)\nzero ,$ and thus, $\epsilon_q =  w_q
- w_0 + 4\nzero v_Z(q)$. Using that $\langle \hat{N} \rangle =
\nzero  + \sum_{\bq\not=0}\langle
b_{\bq}^{\dagger}b_{\bq} \rangle = \nphi /2$, we can calculate
$ \nzero  $, and determine
$\Omega_q$.  The latter is illustrated in
Fig.\ \ref{fig02}(a) for several values of $d/l$.

In the long wavelength limit, $\Omega_q \approx \hbar v q$, where
the linear mode velocity is $v = \sqrt{2n_0\epsilon_B
d/l}(l/\hbar)e^2/\epsilon l$. Here $\epsilon_B =  -d/4l
+\sqrt{\pi/32}(1+d^{2}/l^{2})
\exp\left(d^2/2l^2\right){\rm
Erfc}(d/\sqrt{2}l)$, and ${\rm Erfc}(x)$ is the
complementary error function. For $d/l = 0.2$, $0.5$, $0.8$,
$1.0$, $1.5$, and $2.0$, we find $v = 4.47$, $4.64$, $3.92$,
$3.39$, $2.27$, and $1.52 \times 10^4$ m/s, respectively.

The dispersion relation $\Omega_\bq$ of the quasiparticles
increases with momentum for all values of $d/l$, in qualitative agreement
with calculations done in Refs. \cite{fertig89,macdonald90,joglekar01}
in the region $d/l \le 1.0$.
However, these previous calculations found that a
minimum (roton-like excitation) appears around $|\bq l| \sim 1.0$
for $d/l > 1.0$ and that this minimum becomes soft at $d/l
> 1.2$, whereas, in our case, no minimum arises when $d/l$ increases.
The behavior of the condensate fraction $ n_0 = \nzero /(\nphi /2)$ in terms
of $d/l$ shows that the Bogoliubov approximation for the boson
model (\ref{hboso}) holds only for small $d/l$. As the ratio $d/l$
increases, the number of bosons in the condensate decreases
continuously ($n_0 < 0.3$ for $d/l > 1.0$),
indicating that the interaction between excited (out of
the condensate) bosons starts to play an important role in the
description of the system.

{\it Popov approximation.} It is possible to go a step further in
the analysis of the interacting boson Hamiltonian (\ref{hboso}),
by including the {\it interaction} between the excited bosons, which is
neglected in the Bogoliubov approximation.

Let us consider the four-operator terms of the Hamiltonian
(\ref{hboso}) with $|\bq l|\not= 0$ and treat them on the mean-field
level, including only normal averages. This treatment resembles the
so-called first-order Popov approximation for Bose gases at finite
temperatures \cite{shi98}. By applying this approximation to the
boson model (\ref{hboso}), we obtain a Hamiltonian similar to
Eq.\ (\ref{hboso_bog}) with the replacement
$\epsilon _q \rightarrow \bar{\epsilon}_q =  w_q - \mu +
4\nzero v_Z(0) + 4\nzero v_Z(q) +
4\sum_\mathbf{p\not=q}\left(v_Z(0) + v_{\bp-\bq}(\bq,\bq)\right)
\langle b_{\bp}^{\dagger}b_{\bp}\rangle$. After diagonalizing the
new Hamiltonian and calculating the chemical potential, we find
the following set of self-consistent equations
\beqn
\lambda_q &=& 2\nphi v_Z(q) - 4v_Z(q)\sum_{\bp\not= 0}
\mathcal{F}_p,
\nn \\
\bar{\epsilon}_q &=&  w_q - w_o + 2\nphi v_Z(q)
\nn \\
   &+& 4\sum_{\bp\not= 0} \left[ v_{\bp-\bq}(\bq,\bq) - v_Z(p) -
  v_Z(q)\right]\mathcal{F}_p,
\nn \\
\bar{\Omega}_q &=& \sqrt{\bar{\epsilon}^2_q - \lambda^2_q},
\label{selfequations}  \\
\mathcal{F}_q &=& \left(\bar{\epsilon}_q/\bar{\Omega}_q - 1\right)/2,
\nn \eeqn
where $\bar{\Omega}_q$ is the (new) quasiparticle dispersion relation.

Solutions for the above self-consistent problem can {\it only} be
obtained for 
$d/l \le (d/l)_c = 0.4$. In this case, the dispersion relation of the
quasiparticles also vanishes linearly with momentum for all $d/l$,
qualitatively confirming the results obtained within the
Bogoliubov approximation. The velocities of the linear modes are
renormalized, for instance, $v = 4.30 \times 10^4$ m/s for $d/l =
0.2$.

However, as the ratio $d/l$ increases, it is not possible to find
real solutions for Eqs.\ (\ref{selfequations}), implying that the
ground state of the system is no longer a true condensate of bosons
with zero momentum. This indicates that a new phase sets in above 
$(d/l)_c$ and below the well-known
incompressible-compressible phase transition, which experimentally
takes place at $(d/l)_c^{I-C}$. Notice that $(d/l)_c = 0.4$ is {\it much smaller}
than the critical ratio  $(d/l)_c^{I-C}$, which, in 
samples with negligible electron interlayer tunnelling, is around $1.6-1.8$ 
\cite{kellogg04,wiersma04}, indicating that the novel phase should be
observable within a large region in the parameter space. 
The ground state of the bilayer QHS at $\nu_T = 1$ in the region $(d/l)_c < d/l <
(d/l)_c^{I-C}$ should then be more complex than a pure exciton condensate, possibly  
a zero momentum boson condensate coexisting with a
charge-density wave state as suggested in Ref.\ \cite{chen92}. 
A detailed study of the properties of this novel phase based on the
interacting boson model (\ref{hboso}) and the analysis of the nature
of this new phase transition (or crossover) will be the subject of further
investigations.  

Our findings open up the possibility of the existence of a
two-component phase in bilayers, based solely in an appropriate treatment of
the Coulomb interaction. Disorder \cite{fertig05}, though probably relevant, does not
necessarily need to be invoked in order to generate a richer phase
diagram.   

Finally, we should mention that our findings are quite different from
previous theoretical calculations 
\cite{fertig89, macdonald90, joglekar01} which point out that an
exciton condensate is stable for $d/l \le 1.2$. Within our approach,
this instability appears at even smaller values of $d/l$. Moreover, in
Ref. \cite{macdonald90}, the softness of the roton-like excitation 
at $d/l \sim 1.2$ is associated with the incompressible-compressible
phase transition, which is not the case here. 

{\it Finite electron interlayer tunnelling.} The effect of a
finite electron interlayer tunnelling 
($\Delta_{SAS} \not= 0$) can be easily included in our formalism.
Now the complete Hamiltonian (\ref{total_hamiltoniano}) should be considered. 
It is easy to show that the bosonic representation of $\mathcal{H}_T$
[Eq. (\ref{tunel})] 
is {\it quadratic} in the boson operators. Adding this extra term to the
boson model (\ref{hboso}), and following the same steps as before,
we find,
in the Bogoliubov approximation, that the dispersion relation of the
quasiparticles is also given by Eq.\ (\ref{dispersion}), but now
$\epsilon_q = \Delta_{SAS}(1/4 + 
1/n_0)\sqrt{n_0/2} + w_\bq - w_0 + 4{\rm \nzero }v_Z(q)$. Notice that
the spectrum is no longer gapless. As $\Delta_{SAS}$ increases, a
minimum (roton-like excitation) develops in the excitation spectrum
around $|\bq l| = 2$ for $d/l \ge 1.0$. The energy of the minimum
reduces as the ratio $d/l$ increases. The case $\Delta_{SAS} = 0.1\;
e^2/\epsilon l$ is illustrated in Fig.\ \ref{fig02}(b) for several
values of $d/l$.

\begin{figure}[t]
\centerline{\includegraphics[height=5.0cm]{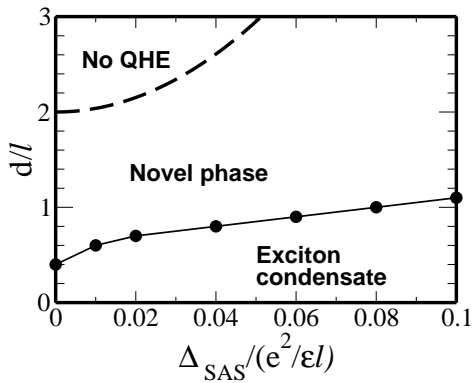}}
\caption{\label{fig03}{Phase diagram ($d/l$ $\times$ $\Delta_{SAS}$)
    of the bilayer QHS at 
    $\nu_T=1$: the filled circles are the calcultated critical ratio $(d/l)_c$ 
    (the solid line is a
    guide to the eyes) and the dashed line is the experimental estimate for
    $(d/l)_c^{I-C}$ (from Ref.\ \cite{murphy94}).}}
\end{figure}

In the Popov approximation, the final set of self-consistent
equations are similar to Eqs. (\ref{selfequations}) with the
replacement $\bar{\epsilon}_q \rightarrow \bar{\bar{\epsilon}}_q =
\bar{\epsilon}_q + \Delta_{SAS}(3n_0 + 2)/\sqrt{32n_0}$. 
The inclusion of a finite eletron interlayer tunnelling increases the
critical ratio $(d/l)_c$ (see Fig.\ \ref{fig03}).
However, the novel phase should remain robust in a sizable region
because $(d/l)_c^{I-C}$ also increases with the
tunnelling strength as verified experimentally \cite{murphy94}.
For $d/l \le (d/l)_c$, the
dispersion relation of the quasiparticles is quite similar to the
results found in the Bogoliubov approximation apart from small
renormalizations. 

{\it Conclusions.} By developing a non-perturbative boso-\\
nization approach for bilayer QHSs at $\nu_T=1$ and by considering
the {\it fully} interacting boson model, we are able to properly
explore the conjecture of a BEC of excitons in the bilayer QHS at
$\nu_T = 1$. 
We show that an exciton condensate is stable only at very small values
of the ratio $d/l$, indicating that the bilayer undergoes a phase
transition (or crossover) at $(d/l)_c$. A new candidate for
the ground state of the system in the region $(d/l)_c < d/l <
(d/l)_c^{I-C}$ is requested because its thorough understanding will be important
for the correct description of the incompressible-compressible phase
transition. Details of the above calculations will be presented
elsewhere.

We would like to thank H.\ T.\ C.\ Stoof, M.\ O.\ Goerbig, E.\
Rezayi, A.\ Muramatsu, and N.\ P.\ Proukakis for helpful discussions. R.L.D. and
A.O.C. kindly acknowledge Funda\c{c}\~ao de Amparo \`a Pesquisa do
Estado de S\~ao Paulo (Fapesp) and Conselho Nacional de
Desenvolvimento Cient\'ifico e Tecnol\'ogico (CNPq) for the
financial support.

\end{document}